\documentclass[journal]{IEEEtran}

\usepackage{amsthm}
\usepackage[fleqn]{amsmath}

\ifCLASSINFOpdf
   \usepackage[pdftex]{graphicx}
\else
   \usepackage[dvips]{graphicx}
   \graphicspath{{../eps/}}
\fi
\usepackage[tight,footnotesize]{subfigure}
\begin{document}
%
\title{Exploring Relay Cooperation for Secure and Reliable Transmission in Two-Hop Wireless Networks}
%
%
%

\author{\IEEEauthorblockN{Yulong Shen\IEEEauthorrefmark{1}\IEEEauthorrefmark{4},
Xiaohong Jiang\IEEEauthorrefmark{2}, Jianfeng
Ma\IEEEauthorrefmark{1} and Weisong
Shi\IEEEauthorrefmark{3}}\\
\IEEEauthorblockA{\IEEEauthorrefmark{1}School of Computer Science
and Technology, Xidian University, China}\\
\IEEEauthorblockA{\IEEEauthorrefmark{2}School of Systems Information
Science, Future University Hakodate, Japan}\\
\IEEEauthorblockA{\IEEEauthorrefmark{3}Department of Computer
Science, Wayne State University, USA}\\
\IEEEauthorblockA{\IEEEauthorrefmark{4}Email:ylshen@mail.xidian.edu.cn
} }

\maketitle

\begin{abstract}

This work considers the problem of secure and reliable information
transmission via relay cooperation in two-hop relay wireless
networks without the information of both eavesdropper channels and
locations. While previous work on this problem mainly studied
infinite networks and their asymptotic behavior and scaling law
results, this papers focuses on a more practical network with finite
number of system nodes and explores the corresponding exact result
on the number of eavesdroppers one network can tolerant to ensure
desired secrecy and reliability. We first study the scenario where
path-loss is equal between all pairs of nodes and consider two
transmission protocols there, one adopts an optimal but complex
relay selection process with less load balance capacity while the
other adopts a random but simple relay selection process with good
load balance capacity. Theoretical analysis is then provided to
determine the maximum number of eavesdroppers one network can
tolerate to ensure a desired performance in terms of the secrecy
outage probability and transmission outage probability. We further
extend our study to the more general scenario where path-loss
between each pair of nodes also depends the distance between them,
for which a new transmission protocol with both preferable relay
selection and good load balance as well as the corresponding
theoretical analysis are presented.

\end{abstract}


%
\IEEEpeerreviewmaketitle

\section{Introduction}

Two-hop ad hoc wireless networks, where each packet travels at most
two hops (source-relay-destination) to reach its destination, have
been a class of basic and important networking scenarios
\cite{IEEEhowto:Sathya}. Actually, the analysis of basic two-hop
relay networks serves as the foundation for performance study of
general multi-hop networks. Due to the promising applications of ad
hoc wireless networks in many important scenarios (like battlefield
networks, vehicle networks, disaster recovery networks), the
consideration of secrecy (and also reliability) in such networks is
of great importance for ensuring the high confidentiality
requirements of these applications.

Traditionally, the information security is provided by adopting the
cryptography approach, where a plain message is encrypted through a
cryptographic algorithm that is hard to break (decrypt) in practice
by any adversary without the key. While the cryptography is
acceptable for general applications with standard security
requirement, it may not be sufficient for applications with a
requirement of strong form of security (like military networks and
emergency networks). This is because that the cryptographic approach
can hardly achieve everlasting secrecy, since the adversary can
record the transmitted messages and try any way to break them
\cite{IEEEhowto:Talbot}. That is why there is an increasing interest
in applying signaling scheme in physical layer to provide a strong
form of security, where a degraded signal at an eavesdropper is
always ensured such that the original data can be hardly recovered
regardless of how the signal is processed at the eavesdropper. We
consider applying physical layer method to achieve secure and
reliable information transmission in the two-hop wireless networks.

By now, a lot of research works have been dedicated to the study of physical
layer security based on cooperative relays and artificial noise, and these works can be roughly classified
into two categories depending on whether the information of eavesdroppers channels
and locations is known or not (see Section V for related works).
 For the case that the information of eavesdroppers channels
and locations is available, a lot of transmission schemes have been
proposed to achieve the maximum secrecy rates while optimizing the
artificial noise generation and power control to reduce the total
transmission power consumption [3-19]. In practice, however, it is
difficult to gain the information of eavesdropper channels and
locations, since the eavesdroppers always try to hide their identity
information as much as possible.
 To alleviate such a requirement on eavesdroppers information, some recent works explored the implementation of secure and reliable information
transmission in wireless networks without the information of both
eavesdropper channels and locations [20-28].
 It is notable, however, that these works mainly focus on exploring the scaling law results in terms of the
number of eavesdroppers one network can tolerate as the number of
system nodes there tends to infinity.
 Although the scaling law results are helpful for us to
understand the general asymptotic network behavior,  they tell us
a little about the actual and exact number of eavesdroppers one network can tolerate. In practice, however, such exact
results are of great interest for network designers.


This paper focuses on applying
the relay cooperation to achieve secure and reliable information
transmission in a more practical finite two-hop wireless network
without the knowledge of both eavesdropper channels and locations.
The main contributions of this paper as follows.

\begin{itemize}

\item
For achieving secure and reliable information transmission in a more
practical two-hop wireless network with finite number of system
nodes and equal path-loss between all pairs of
nodes, we consider the application of the cooperative protocol
proposed in \cite{IEEEhowto:Goeckel2} with an optimal and complex
relay selection process but less load balance capacity, and also
propose to use a new cooperative protocol with a simple and random
relay selection process but good load balance capacity.

\item
Rather than exploring the asymptotic behavior and scaling law
results, we provide theoretic analysis for above two cooperative protocols to determine the corresponding
exact results on the number of eavesdroppers one network can
tolerate to meet a specified requirement in terms of the maximum
secrecy outage probability and the maximum transmission outage
probability allowed.

\item

We further extend our study to the more general and practical scenario where
the path-loss between each pair of nodes also depends on their relative
locations, for which we propose a new transmission protocol with both preferable relay
selection and good load balance and also present the corresponding theoretical analysis under this new protocol.


\end{itemize}

The remainder of the paper is organized as follows. Section II
presents system models and also introduces transmission outage and
secrecy outage for the analysis of transmission protocols.
Section III considers two transmission protocols for the scenario of equal
path-loss between all pairs of nodes and provides the corresponding theoretical
analysis. Section IV further presents a new transmission protocol and its theoretical
analysis to address distance-dependent path-loss issue. Section V introduces the related works and Section VI concludes this paper.

\section{System Models}

\subsection{Network Model}
As illustrated in Fig.1 that we consider a network scenario where a
source node $S$ wishes to communicate securely with its destination
node $D$ with the help of multiple relay nodes $R_1$, $R_2$,
$\cdots$, $R_n$. In addition to these normal system nodes, there are
also $m$ eavesdroppers $E_1$, $E_2$, $\cdots$, $E_m$ that are
independent and also uniformly distributed in the network. Our goal
here is to ensure the secure and reliable information transmission
from source $S$ to destination $D$ under the condition that no real
time information is available about both eavesdropper channels and
locations.

\begin{figure}[!t]
\centering
\includegraphics[width=2in]{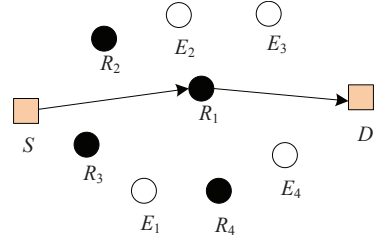}
\DeclareGraphicsExtensions. \caption{System scenario: Source $S$
wishes to communicate securely with destination $D$ with the
assistance of finite relays $R_1$, $R_2$, $\cdots$, $R_{n}$ ($n$=4
in the figure) in the presence of passive eavesdroppers $E_1$,
$E_2$, $\cdots$, $E_{m}$ ($m$=4 in the figure). Cooperative relay
scheme is used in the two-hop transmission. A assistant node is
selected randomly as relay ($R_1$ in the figure).} \label{System
scenario}
\end{figure}

\subsection{Transmission Model}

Consider the transmission from a transmitter $A$ to a receiver $B$,
and denote by $x_i^{\left(A\right)}$ the $i^{th}$ symbol transmitted
by $A$ and denote by $y_i^{\left(B\right)}$ the $i^{th}$ signal
received by $B$. We assume that all nodes transmit with the same
power $E_s$, path-loss between all pairs of nodes is independent,
and the frequency-nonselective multi-path fading from $A$ to $B$ is
a complex zero-mean Gaussian random variable. Under the condition
that all nodes in a group of nodes, $\mathcal {R}$, are generating
noises, the $i^{th}$ signal received at node $B$ from node $A$ is
determined as:

$$y_i^{\left(B\right)}=\frac{h_{A,B}}{d_{A,B}^{\alpha / 2}} \sqrt{E_s}x_i^{\left(A\right)} +
\sum_{A_i \in \mathcal {R}} \frac{h_{A_i,B}}{d_{A_i,B}^{\alpha / 2}}
\sqrt{E_s}x_i^{\left(A_i\right)} + n_i^{\left(B\right)}$$

where $\alpha \geq 2$ is the path-loss exponent. The noise
$\left\{n_i^{\left(B\right)}\right\}$ at receiver $B$ is assumed to
be i.i.d complex Gaussian random variables with
$E{\left[\left|n_i^{\left(B\right)}\right|^2\right]} = N_0$, and
$\left|h_{A,B}\right|^2$ is exponentially distributed with mean
$E{\left[\left|h_{A,B}\right|^2\right]}$. Without loss of
generality, we assume that
$E{\left[\left|h_{A,B}\right|^2\right]}=1$. The SINR $C_{A,B}$ from
$A$ to $B$ is then given by

$$C_{A,B}=\frac{E_s\left|h_{A,B}\right|^2 d_{A,B}^{- \alpha}}{\sum_{A_i \in \mathcal
{R}}E_s{\left|h_{A_i,B}\right|^2 d_{A_i,B}^{- \alpha}}+N_0/2}$$

For a legitimate node and an eavesdropper, we use two separate SINR
thresholds $\gamma_R$ and $\gamma_E$ to define the minimum SINR
required to recover the transmitted messages for legitimate node and
eavesdropper, respectively. Therefore, a system node (relay or
destination) is able to decode a packet if and only if its SINR is
greater than $\gamma_R$, while the transmitted message is secure if
and only if the SINR at each eavesdropper is less than $\gamma_E$.

\subsection{Transmission Outage and Secrecy Outage}

For a transmission from the source $S$ to destination $D$, we call
transmission outage happens if $D$ can not decode the transmitted
packet, i.e., $D$ received the packet with SINR less than the
predefined threshold $\gamma_R$. The transmission outage
probability, denoted as $P_{out}^{\left(T\right)}$, is then defined
as the probability that transmission outage from $S$ to $D$ happens.
For a predefined upper bound $\varepsilon_t$ on
$P_{out}^{\left(T\right)}$, we call the communication between $S$
and $D$ is reliable if $P_{out}^{\left(T\right)} \leq
\varepsilon_t$. Notice that for the transmissions from $S$ to the
selected relay $R_{j^\ast}$ and from $R_{j^\ast}$ to $D$, the
corresponding transmission outage can be defined in the similar way
as that of from $S$ to $D$. We use $O_{S \rightarrow
R_{j^\ast}}^{(T)}$ and $O_{R_{j^\ast} \rightarrow D}^{(T)}$ to
denote the events that transmission outage from source $S$ to
$R_{j^\ast}$ happens and transmission outage from relay $R_{j^\ast}$
to $D$ happens, respectively. Due to the link independence
assumption, we have

\begin{align*}
&P_{out}^{\left(T\right)} =P\left(O_{S \rightarrow R_{j^\ast}}^{(T)}
\cup O_{R_{j^\ast} \rightarrow D}^{(T)}\right)\\
&\ \ \ \ \ \ = P\left(O_{S \rightarrow
R_{j^\ast}}^{(T)}\right)+P\left(O_{R_{j^\ast}
\rightarrow D}^{(T)}\right)\\
&\ \ \ \ \ \ \ \ \ -P\left(O_{S \rightarrow R_{j^\ast}}^{(T)}\right)
\cdot P\left(O_{R_{j^\ast} \rightarrow D}^{(T)}\right)
\end{align*}

Regarding the secrecy outage, we call secrecy outage happens for a
transmission from $S$ to $D$ if at least one eavesdropper can
recover the transmitted packets during the process of this two-hop
transmission, i.e., at least one eavesdropper received the packet
with SINR larger than the predefined threshold $\gamma_E$. The
secrecy outage probability, denoted as $P_{out}^{\left(S\right)}$,
is then defined as the probability that secrecy outage happens
during the transmission from $S$ to $D$. For a predefined upper
bound $\varepsilon_s$ on $P_{out}^{\left(S\right)}$, we call the
communication between $S$ and $D$ is secure if
$P_{out}^{\left(S\right)} \leq \varepsilon_s$. Notice that for the
transmissions from $S$ to the selected relay $R_{j^\ast}$ and from
$R_{j^\ast}$ to $D$, the corresponding secrecy outage can be defined
in the similar way as that of from $S$ to $D$. We use $O_{S
\rightarrow R_{j^\ast}}^{(S)}$ and $O_{R_{j^\ast} \rightarrow
D}^{(S)}$ to denote the events that secrecy outage from source $S$
to $R_{j^\ast}$ happens and secrecy outage from relay $R_{j^\ast}$
to $D$ happens, respectively. Again, due to the link independence
assumption, we have

\begin{align*}
&P_{out}^{\left(S\right)} =P\left(O_{S \rightarrow
R_{j^\ast}}^{(S)}\right)+P\left(O_{R_{j^\ast} \rightarrow
D}^{(S)}\right)\\
&\ \ \ \ \ \ \ \ \ -P\left(O_{S \rightarrow R_{j^\ast}}^{(S)}\right)
\cdot P\left(O_{R_{j^\ast} \rightarrow D}^{(S)}\right)
\end{align*}

\section{Secure and Reliable Transmission under Equal Path-Loss}

In this section, we consider the case where the path-loss is equal
between all pairs of nodes in the system (i.e., we set $d_{A,B} = 1$ for all
$A \neq B$). We first introduce two transmission protocols
considered for such scenario, and then provide theoretical analysis to
determine the numbers of eavesdroppers one network can tolerate
under these protocols.

\subsection{Transmission Protocols}
The first protocol we consider (hereafter called Protocol 1) is the one proposed in
\cite{IEEEhowto:Goeckel2}, in which the optimal relay node with the
best link condition to both source and destination is always
selected for information relaying.
Notice that although the Protocol 1 can guarantee the optimal relay node selection, it suffers from several problems. Protocol
1 involves a complicated process of optimal relay selection, which
is not very suitable for the distributed wireless networks, in
particular when the number of possible relay nodes is huge. More
importantly, since the channel state is relatively constant during a
fixed time period, some relay nodes with good link conditions are
always preferred for information relaying, resulting in a severe
load balance problem and a quick node energy depletion in
energy-limited wireless environment.

Based on above observations, we propose to use a simple and random
relay selection rather than the optimal relay selection in Protocol 1 to achieve a
better load and energy consumption balance among possible relay
nodes. By modifying the Protocol 1, the new transmission protocol (hereafter called Protocol 2)
works as follows.

\textbf{1) \emph{Relay selection}:} A relay node, indexed by
$j^\ast$, is randomly selected from all candidate relay nodes $R_j,
j=1,2,\cdots,n$.

\textbf{2) \emph{Channel measurement}:} The selected relay
$R_{j^\ast}$ broadcasts a pilot signal to allow each of other relays
to measure the channel from $R_{j^\ast}$ to itself. Each of the
other relays $R_j, j=1,2,\cdots,n, j \neq j^\ast$ then knows the
corresponding value of $h_{R_j,R_{j^\ast}}$. Similarly, the
destination $D$ broadcasts a pilot signal to allow each of other
relays to measure the channel from $D$ to itself. Each of the other
relays $R_j, j=1,2,\cdots,n, j \neq j^\ast$ then knows the
corresponding value of $h_{R_j,D}$.

\textbf{3) \emph{Message transmission}:} The source $S$ transmits the messages
to $R_{j^\ast}$, and concurrently, the relay nodes with indexes in
$\mathcal {R}_1 = {\left\{j \neq j^\ast : |h_{R_j,R_{j^\ast}}|^2 <
\tau \right\}}$ transmit noise to generate interference at
eavesdroppers. The relay
$R_{j^\ast}$ then transmits the messages to destination $D$, and concurrently, the relay nodes with indexes in $\mathcal {R}_2 =
{\left\{j \neq j^\ast : |h_{R_j,D}|^2 < \tau \right\}}$ transmit
noise to generate interference at eavesdroppers.

\emph{Remark 1}: The parameter $\tau$ involved in the Protocol 1 and
Protocol 2 serves as the threshold on path-loss, based on which the set of noise generating relay nodes can be identified. Notice that a too large $\tau$ may disable legitimate transmission, while a too small $\tau$
may not be sufficient for interrupting all eavesdroppers. Thus, the parameter $\tau$ should be set properly to ensure both secrecy requirement and reliability requirement.

\emph{Remark 2}: The two protocols considered here have their own
advantages and disadvantages and thus are suitable for different
network scenarios. For the protocol 1, it
can achieve a better performance in terms of the number of
eavesdroppers can be tolerated (see Theorem 1). However, it involves a complex
relay selection process, and more importantly, it results in an
unbalanced load and energy consumption distribution among
systems nodes. Thus, such protocol is suitable for small
scale wireless network with sufficient energy supply rather
than large and energy-limited wireless networks (like wireless
sensor networks). Regarding the Protocol 2, although it can
tolerate less number eavesdroppers in comparison with the
Protocol 1 (see Theorem 2), it involves a very simple random relay selection
process to achieve a good load and energy consumption
distribution among system nodes. Thus, this protocol is more
suitable for large scale wireless network environment with
stringent energy consumption constraint.

\subsection{Analysis of Protocol 1}

We now analyze that under the Protocol 1 the number of eavesdroppers one network can
tolerate subject to specified requirements on transmission outage and
secrecy outage. We first establish the following two lemmas regarding some basic properties of $P_{out}^{\left(T\right)}$, $P_{out}^{\left(S\right)}$ and $\tau$, which will help us to derive the main result in Theorem 1.

\emph{Lemma 1}: Consider the network scenario of Fig 1 with equal
path-loss between all pairs of nodes, under the Protocol 1 the
transmission outage probability $P_{out}^{\left(T\right)}$ and
secrecy outage probability $P_{out}^{\left(S\right)}$ there satisfy
the following conditions.

\begin{align*}
&P_{out}^{\left(T\right)} \leq
2\left[1-e^{-2\gamma_R\left(n-1\right)\left(1-e^{-\tau}\right)\tau}\right]^n\\
&\ \ \ \ \ \ \ \ \
-\left[1-e^{-2\gamma_R\left(n-1\right)\left(1-e^{-\tau}\right)\tau}\right]^{2n}
\end{align*}
\begin{align*}
&P_{out}^{\left(S\right)} \leq 2m \cdot \left(\frac{1}{1+\gamma_E}\right)^{\left(n-1\right)\left(1-e^{-\tau}\right)}\\
& \ \ \ \ \ \ \ \ \ \ -\left[m \cdot
\left(\frac{1}{1+\gamma_E}\right)^{\left(n-1\right)\left(1-e^{-\tau}\right)}\right]^2
\end{align*}

The proof of Lemma 1 can be found in the Appendix A.

\emph{Lemma 2}: Consider the network scenario of Fig 1 with equal
path-loss between all pairs of nodes, to ensure
$P_{out}^{\left(T\right)} \leq \varepsilon_t$ and
$P_{out}^{\left(S\right)} \leq \varepsilon_s$ under the Protocol 1,
the parameter $\tau$ must satisfy the following condition.

\begin{align*}
\tau \leq \sqrt{\frac{-\log\left[1- \left(1 - \sqrt{1 -
\varepsilon_t}\right)^{\frac{1}{n}}\right]}{2\gamma_R\left(n-1\right)}}
\end{align*}

and

\begin{align*}
& \tau \geq - \log{\left[1 + \frac{\log{\left(\frac{1 - \sqrt{1 -
\varepsilon_s}}{m}\right)}}{\left(n - 1\right)\log{\left(1 +
\gamma_E\right)}}\right]}
\end{align*}

\begin{proof}

\textbf{$\bullet$ Reliability Guarantee}

To ensure the reliability requirement $P_{out}^{\left(T\right)}
\leq \varepsilon_t$, we know from the Lemma 1 that we just need

\begin{align*}
&2\left[1-e^{-2\gamma_R\left(n-1\right)\left(1-e^{-\tau}\right)\tau}\right]^n\\
&-\left[1-e^{-2\gamma_R\left(n-1\right)\left(1-e^{-\tau}\right)\tau}\right]^{2n}\\
&\leq \varepsilon_t
\end{align*}

Thus,

$$\left[1-e^{-2\gamma_R\left(n-1\right)\left(1-e^{-\tau}\right)\tau}\right]^n \leq 1 - \sqrt{1 - \varepsilon_t}$$

That is,

$$-2\gamma_R\left(n-1\right)
\left(1-e^{-\tau}\right)\tau \geq \log\left[1- \left(1 - \sqrt{1 -
\varepsilon_t}\right)^{\frac{1}{n}}\right]$$

By using Taylor formula, we have

\begin{align*}
\tau \leq \sqrt{\frac{-\log\left[1- \left(1 - \sqrt{1 -
\varepsilon_t}\right)^{\frac{1}{n}}\right]}{2\gamma_R\left(n-1\right)}}
\end{align*}

The above result indicates the maximum value the parameter $\tau$ we can take
to ensure the reliability requirement.

\textbf{$\bullet$ Secrecy Guarantee}

To ensure the secrecy requirement $P_{out}^{\left(S\right)} \leq
\varepsilon_s$ , we know from the Lemma 1 that we just need

\begin{align*}
&2m \cdot \left(\frac{1}{1+\gamma_E}\right)^{\left(n-1\right)\left(1-e^{-\tau}\right)}\\
& -\left[m \cdot
\left(\frac{1}{1+\gamma_E}\right)^{\left(n-1\right)\left(1-e^{-\tau}\right)}\right]^2\\
&\leq \varepsilon_s
\end{align*}

Thus,

\begin{align*}
&m \cdot
\left(\frac{1}{1+\gamma_E}\right)^{\left(n-1\right)\left(1-e^{-\tau}\right)}
\leq 1- \sqrt{1-\varepsilon_s}
\end{align*}

That is,

\begin{align*}
& \tau \geq - \log{\left[1 + \frac{\log{\left(\frac{1 - \sqrt{1 -
\varepsilon_s}}{m}\right)}}{\left(n - 1\right)\log{\left(1 +
\gamma_E\right)}}\right]}
\end{align*}

The above result implies the minimum value parameter $\tau$ we can take to
guarantee the secrecy requirement.

\end{proof}

Based on the results of Lemma 2, we now can establish the following
theorem regarding the performance of Protocol 1.

\textbf{Theorem 1.} Consider the network scenario of Fig 1 with
equal path-loss between all pairs of nodes. To guarantee
$P_{out}^{\left(T\right)} \leq \varepsilon_t$ and
$P_{out}^{\left(S\right)} \leq \varepsilon_s$ under the Protocol 1,
the number of eavesdroppers $m$ one network can tolerate must
satisfy the following condition.

\begin{align*}
& m \leq \left(1 - \sqrt{1 - \varepsilon_s}\right) \cdot
\left(1+\gamma_E\right)^{\sqrt{\frac{-\left(n-1\right)\log\left[1-
\left(1 - \sqrt{1 -
\varepsilon_t}\right)^{\frac{1}{n}}\right]}{2\gamma_R}}}
\end{align*}

\begin{proof}

From Lemma 2, we know that to ensure the reliability requirement, we
have

\begin{align*}
\tau \leq \sqrt{\frac{-\log\left[1- \left(1 - \sqrt{1 -
\varepsilon_t}\right)^{\frac{1}{n}}\right]}{2\gamma_R\left(n-1\right)}}
\end{align*}

and

\begin{align*}
\left(n-1\right) \left(1-e^{-\tau}\right) \leq \frac{-\log\left[1-
\left(1 - \sqrt{1 -
\varepsilon_t}\right)^{\frac{1}{n}}\right]}{2\gamma_R\tau}
\end{align*}

To ensure the secrecy requirement, we need

\begin{align*}
& \left(\frac{1}{1+\gamma_E}\right)^{\left(n
-1\right)\left(1-e^{-\tau}\right)} \leq \frac{1 - \sqrt{1 -
\varepsilon_s}}{m}
\end{align*}

Thus,

\begin{align*}
& m \leq \frac{1 - \sqrt{1 -
\varepsilon_s}}{\left(\frac{1}{1+\gamma_E}\right)^{\left(n
-1\right)\left(1-e^{-\tau}\right)}}\\
&\ \ \ \ \leq \frac{1 - \sqrt{1 -
\varepsilon_s}}{\left(\frac{1}{1+\gamma_E}\right)^{\frac{-\log\left[1-
\left(1 - \sqrt{1 -
\varepsilon_t}\right)^{\frac{1}{n}}\right]}{2\gamma_R\tau}}}
\end{align*}

By letting $\tau$ to take its maximum value for maximum interference at
eavesdroppers, we get the following bound

\begin{align*}
& m \leq \left(1 - \sqrt{1 - \varepsilon_s}\right) \cdot
\left(1+\gamma_E\right)^{\sqrt{\frac{-\left(n-1\right)\log\left[1-
\left(1 - \sqrt{1 -
\varepsilon_t}\right)^{\frac{1}{n}}\right]}{2\gamma_R}}}
\end{align*}

\end{proof}

\subsection{Analysis of Protocol 2}

Similar to the analysis of Protocol 1, we first establish the
following two lemmas regarding some basic properties of
$P_{out}^{\left(T\right)}$, $P_{out}^{\left(S\right)}$ and $\tau$
under the Protocol 2.

\emph{Lemma 3}: Consider the network scenario of Fig 1 with equal
path-loss between all pairs of nodes, the transmission outage
probability $P_{out}^{\left(T\right)}$ and secrecy outage
probability $P_{out}^{\left(S\right)}$ under the Protocol 2 satisfy
the following conditions.

\begin{align*}
&P_{out}^{\left(T\right)} \leq 2\left[1 -
e^{-\gamma_R\left(n-1\right)\left(1-e^{-\tau}\right)\tau}\right]
\\
&\ \ \ \ \ \ \ \ \ -\left[1 -
e^{-\gamma_R\left(n-1\right)\left(1-e^{-\tau}\right)\tau}\right]^2
\end{align*}

\begin{align*}
&P_{out}^{\left(S\right)} \leq 2m \cdot \left(\frac{1}{1+\gamma_E}\right)^{\left(n-1\right)\left(1-e^{-\tau}\right)}\\
& \ \ \ \ \ \ \ \ \ \ -\left[m \cdot
\left(\frac{1}{1+\gamma_E}\right)^{\left(n-1\right)\left(1-e^{-\tau}\right)}\right]^2
\end{align*}

The proof of Lemma 3 can be found in the Appendix B.

\emph{Lemma 4}: Consider the network scenario of Fig 1 with equal
path-loss between all pairs of nodes, to ensure
$P_{out}^{\left(T\right)} \leq \varepsilon_t$ and
$P_{out}^{\left(S\right)} \leq \varepsilon_s$ under the Protocol 2,
the parameter $\tau$ must satisfy the following condition.

$$\tau \in \left[- \log{\left[1 + \frac{\log{\left(\frac{1 - \sqrt{1 -
\varepsilon_s}}{m}\right)}}{\left(n - 1\right)\log{\left(1 +
\gamma_E\right)}}\right]},
\sqrt{\frac{-\log\left(1-\varepsilon_t\right)}{2\gamma_R \left(n
-1\right)}}\right]$$

\begin{proof}

\textbf{$\bullet$ Reliability Guarantee}

To ensure the reliability requirement $P_{out}^{\left(T\right)} \leq
\varepsilon_t$, we know from Lemma 4 that we just need

\begin{align*}
& 2\left[1 -
e^{-\gamma_R\left(n-1\right)\left(1-e^{-\tau}\right)\tau}\right]
\\
& -\left[1 -
e^{-\gamma_R\left(n-1\right)\left(1-e^{-\tau}\right)\tau}\right]^2\\
& \leq \varepsilon_t
\end{align*}

That is,

\begin{align*}
& 1 - e^{-\gamma_R\left(n -1\right)\left(1-e^{-\tau}\right)\tau}
\leq 1 - \sqrt{1 - \varepsilon_t}
\end{align*}

By using Taylor formula, we have

\begin{align*}
\tau \leq \sqrt{\frac{-\log\left(1-\varepsilon_t\right)}{2\gamma_R
\left(n -1\right)}}
\end{align*}

\textbf{$\bullet$ Secrecy Guarantee}

Notice that the secrecy outage probability of Protocol 1 and
Protocol 2 is same. Thus, to ensure the secrecy requirement, we need

\begin{align*}
& \left(\frac{1}{1+\gamma_E}\right)^{\left(n
-1\right)\left(1-e^{-\tau}\right)} \leq \frac{1 - \sqrt{1 -
\varepsilon_s}}{m}
\end{align*}

Thus,

\begin{align*}
& \tau \geq - \log{\left[1 + \frac{\log{\left(\frac{1 - \sqrt{1 -
\varepsilon_s}}{m}\right)}}{\left(n - 1\right)\log{\left(1 +
\gamma_E\right)}}\right]}
\end{align*}

The above result implies the minimum value parameter $\tau$ can take
to guarantee the secrecy requirement.

\end{proof}

\textbf{Theorem 2.} Consider the network scenario of Fig 1 with
equal path-loss between all pairs of nodes. To guarantee
$P_{out}^{\left(T\right)} \leq \varepsilon_t$ and
$P_{out}^{\left(S\right)} \leq \varepsilon_s$ based on the Protocol
2, the number of eavesdroppers $m$ the network can tolerate must
satisfy the following condition.

\begin{align*}
& m \leq \left(1 - \sqrt{1 -
\varepsilon_s}\right)\cdot\left(1+\gamma_E\right)^{\sqrt{\frac{-\left(n-1\right)\log\left(1-\varepsilon_t\right)}{2\gamma_R}}}
\end{align*}

\begin{proof}

From Lemma 4, we know that to ensure the reliability requirement, we
have

\begin{align*}
\tau \leq \sqrt{\frac{-\log\left(1-\varepsilon_t\right)}{2\gamma_R
\left(n -1\right)}}
\end{align*}

and

\begin{align*}
& \left(n - 1\right)\left(1 - e^{-\tau}\right) \leq
\frac{-\log\left(1-\varepsilon_t\right)}{2\gamma_R \tau}
\end{align*}

To ensure the secrecy requirement, we need

\begin{align*}
& \left(\frac{1}{1+\gamma_E}\right)^{\left(n
-1\right)\left(1-e^{-\tau}\right)} \leq \frac{1 - \sqrt{1 -
\varepsilon_s}}{m}
\end{align*}

Thus,

\begin{align*}
& m \leq \frac{1 - \sqrt{1 -
\varepsilon_s}}{\left(\frac{1}{1+\gamma_E}\right)^{\left(n
-1\right)\left(1-e^{-\tau}\right)}}\\
&\ \ \ \leq \frac{1 - \sqrt{1 -
\varepsilon_s}}{\left(\frac{1}{1+\gamma_E}\right)^{\frac{-\log\left(1-\varepsilon_t\right)}{2\gamma_R
\tau}}}
\end{align*}

By letting $\tau$ to take its maximum value for maximum interference at
eavesdroppers, we get the following bound

\begin{align*}
& m \leq \left(1 - \sqrt{1 -
\varepsilon_s}\right)\cdot\left(1+\gamma_E\right)^{\sqrt{\frac{-\left(n-1\right)\log\left(1-\varepsilon_t\right)}{2\gamma_R}}}
\end{align*}

\end{proof}

\section{Secure and Reliable Transmission under Distance-Dependent Path-Loss}

In this section, we consider the more general scenario where the path-loss between
each pair of nodes also depends the distance between them. We first introduce the coordinate system adopted in our discussion, and then propose a flexible transmission protocol to achieve both the preferable relay selection
and good load balance under such distance-dependent path-loss scenario. The related theoretic analysis is further provided to
determine the number of eavesdroppers one network can tolerate by adopting this protocol.

\subsection{Coordinate System}

To address the distance-dependent path-loss, we consider a two-hop relay wireless network deployed in a square of unit area and defined by the the coordinate system shown in Fig.2, where the source $S$ located at coordinate $\left(0,0.5\right)$ wishes to establish two-hop transmission
with destination $D$ located at coordinate $\left(1,0.5\right)$. In addition to the source $S$ and destination $D$, we assume that  there are $n$ cooperative relays and $m$ eavesdroppers of unknown channels and locations  independently and uniformly distributed in the network area.

\begin{figure}[!t]
\centering
\includegraphics[width=3.7in]{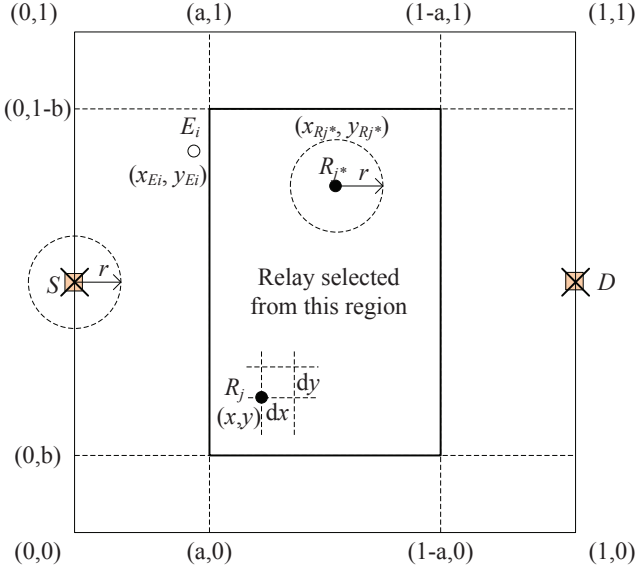}
\DeclareGraphicsExtensions. \caption{Coordinate system for the
scenario where path-loss between pairs of nodes is based on their
relative locations.} \label{Coordinate system}
\end{figure}

\subsection{Transmission Protocol}

Notice that under the distance-dependent path-loss scenario, the
further the distance between a transmitter and a receiver, the
weaker the signal received at the receiver. Thus, the system nodes
located in the middle region between source $S$ and destination
$D$ are preferable relays. Based such observation, we propose here a
general and practical protocol (hereafter called Protocol 3) to
ensure both the preferable relay selection and good load balance for
distance-dependent path-loss scenario, which works as follow.

\textbf{1) \emph{Relay selection}:} Based on two parameters $a$ and $b$, $0 \leq a \leq 0.5, 0
\leq b \leq 0.5$, we first define a relay
selection region
$\left[a, 1-a\right] \times \left[b, 1-b\right]$ between source $S$ and destination $D$. A relay node, indexed
by $j^\ast$, is then selected randomly from relays falling within the
relay selection region.

\textbf{2) \emph{Channel measurement}:} Each of the other relays
measures the channel from the selected relay $R_{j^\ast}$ and
destination $D$ by accepting the pilot signal from $R_{j^\ast}$ and
$D$ for determining the noise generation nodes.

\textbf{3) \emph{Two-hop transmission}:} The source $S$ and
the selected relay $R_{j^\ast}$ transmit the messages in two-hop
transmission. Concurrently, the relay nodes with indexes in
$\mathcal {R}_1 = {\left\{j \neq j^\ast : |h_{R_j,R_{j^\ast}}|^2 <
\tau \right\}}$ in the first hop and the relay nodes with indexes in
$\mathcal {R}_2 = {\left\{j \neq j^\ast : |h_{R_j,D}|^2 < \tau
\right\}}$ in the second hop transmit noise respectively to help
transmission.

\emph{Remark 4}: In the Protocol 3, a trade off between
the preferable relay selection and better load balance can be controlled through the parameters
$a$ and $b$, which define the relay selection region. As to be shown in Theorem 3 that by adopting a small value for both $a$ and $b$ (i.e., a larger relay selection region), a better load balance
capacity can be achieved at the cost of a smaller number of eavesdroppers
one network can tolerant.

\subsection{Analysis of Protocol 3}
To address the near eavesdropper problem and also to simply the analysis for the Protocol 3, we assume that there exits a constant $r_0>0$ such that any eavesdropper falling within a circle area with radius $r_0$ and center $S$ or $R_{j^\ast}$ can eavesdrop the transmitted
messages successfully with probability 1, while any eavesdropper beyond such area can only successfully eavesdropper the transmitted
messages with a probability less than 1. Based on such a simplification, we can establish the following two
lemmas regarding some basic properties of
$P_{out}^{\left(T\right)}$, $P_{out}^{\left(S\right)}$ and $\tau$
under this protocol.

\emph{Lemma 5}: Consider the network scenario of Fig 2, under the
Protocol 3 the transmission outage probability
$P_{out}^{\left(T\right)}$ and secrecy outage probability
$P_{out}^{\left(S\right)}$ there satisfy the following conditions.

\begin{equation*}
\begin{aligned}
&P_{out}^{\left(T\right)} \leq \left[1-e^{-\frac{\gamma_R {\tau
\left(n-1\right)\left(1 -
e^{-\tau}\right)}}{\phi^{-\alpha}}\left(\varphi_1+\varphi_2\right)}\right]\left(1
- \vartheta\right)+ 1 \cdot \vartheta
\end{aligned}
\end{equation*}

\begin{equation*}
\begin{aligned}
&P_{out}^{\left(S\right)} \leq 2 m \left[\pi {r_0}^2 +
\left(\frac{1}{1 + \gamma_E \psi
{r_0}^{\alpha}}\right)^{\left(n-1\right)\left(1-e^{-\tau}\right)}\left(1- \pi {r_0}^2\right)\right]\\
&\ \ -\left[m \left(\pi {r_0}^2 + \left(\frac{1}{1 + \gamma_E \psi
{r_0}^{\alpha}}\right)^{\left(n-1\right)\left(1-e^{-\tau}\right)}\left(1-
\pi {r_0}^2\right)\right)\right]^2
\end{aligned}
\end{equation*}

here,

\begin{align*}
&\vartheta = \bigg[1-\left(1-2a\right)\left(1-2b\right)\bigg]^n
\end{align*}

\begin{align*}
&\varphi_1 =  \int_0^1\int_0^1\frac{1}{\left[\left(x - 0.5\right)^2
+ \left(y - 0.5\right)^2\right]^{\frac{\alpha}{2}}}dxdy
\end{align*}

\begin{align*}
&\varphi_2 =  \int_0^1\int_0^1\frac{1}{\left[\left(x - 1\right)^2 +
\left(y - 0.5\right)^2\right]^{\frac{\alpha}{2}}}dxdy
\end{align*}

\begin{align*}
&\phi = \sqrt{(1-a)^2+(0.5-b)^2}
\end{align*}

\begin{align*}
&\psi = \int_0^1\int_0^1\frac{1}{\left(x ^2 +
y^2\right)^{\frac{\alpha}{2}}}dxdy
\end{align*}

The proof of the Lemma 5 can be found in the Appendix C.

\emph{Lemma 6}: Consider the network scenario of Fig 2, to ensure
$P_{out}^{\left(T\right)} \leq \varepsilon_t$ and
$P_{out}^{\left(S\right)} \leq \varepsilon_s$ by applying the Protocol
3, the parameter $\tau$ must satisfy the following condition.

\begin{align*}
\tau \leq \sqrt{\frac{-\log\left(\frac{1-\varepsilon_t}{1 -
\vartheta}\right){\phi^{-\alpha}}}{\gamma_R
{\left(n-1\right)\left(\varphi_1+\varphi_2\right)}}}
\end{align*}

and

\begin{equation*}
\tau \geq - \log\left[1 + \frac{\log{\left(\frac{\frac{1 - \sqrt{1 -
\varepsilon_s}}{m} - \pi {r_0}^2}{1-\pi
{r_0}^2}\right)}}{\left(n-1\right)\log{\left(1 + \gamma_E \psi
{r_0}^{\alpha}\right)}}\right]
\end{equation*}

here, $\vartheta$, $\varphi_1$, $\varphi_2$, $\phi$ and $\psi$ are
defined in the same way as that in Lemma 5.

\begin{proof}

\textbf{$\bullet$ Reliability Guarantee}

To ensure the reliability requirement $P_{out}^{\left(T\right)} \leq
\varepsilon_t$, we know from Lemma 5 that we just need

\begin{equation*}
\begin{aligned}
&\left[1-e^{-\frac{\gamma_R {\tau \left(n-1\right)\left(1 -
e^{-\tau}\right)}}{\phi^{-\alpha}}\left(\varphi_1+\varphi_2\right)}\right]\left(1
- \vartheta\right)+ 1 \cdot \vartheta \leq \varepsilon_t
\end{aligned}
\end{equation*}

that is,

$$-\frac{\gamma_R {\tau
\left(n-1\right)\left(1 -
e^{-\tau}\right)}}{\phi^{-\alpha}}\left(\varphi_1+\varphi_2\right)
\geq \log{\left(\frac{1-\varepsilon_t}{1 - \vartheta}\right)}$$

By using Taylor formula, we have

\begin{align*}
\tau \leq \sqrt{\frac{-\log\left(\frac{1-\varepsilon_t}{1 -
\vartheta}\right){\phi^{-\alpha}}}{\gamma_R
{\left(n-1\right)\left(\varphi_1+\varphi_2\right)}}}
\end{align*}

\textbf{$\bullet$ Secrecy Guarantee}

To ensure the secrecy requirement $P_{out}^{\left(S\right)} \leq
\varepsilon_s$, we know from Lemma 5 that we just need

\begin{equation*}
\begin{aligned}
&2 m \left[\pi {r_0}^2 + \left(\frac{1}{1 + \gamma_E \psi
{r_0}^{\alpha}}\right)^{\left(n-1\right)\left(1-e^{-\tau}\right)}\left(1-
\pi {r_0}^2\right)\right]-\\
&\left[m \left(\pi {r_0}^2 + \left(\frac{1}{1 + \gamma_E \psi
{r_0}^{\alpha}}\right)^{\left(n-1\right)\left(1-e^{-\tau}\right)}\left(1-
\pi {r_0}^2\right)\right)\right]^2 \\
& \leq \varepsilon_s
\end{aligned}
\end{equation*}

Thus,

\begin{equation*}
\begin{aligned}
& m \cdot \left[\pi {r_0}^2 + \left(\frac{1}{1 + \gamma_E \psi
{r_0}^{\alpha}}\right)^{\left(n-1\right)\left(1-e^{-\tau}\right)} \left(1- \pi {r_0}^2\right) \right]\\
& \leq 1- \sqrt{1-\varepsilon_s}
\end{aligned}
\end{equation*}

That is,

\begin{equation*}
\tau \geq - \log\left[1 + \frac{\log{\left(\frac{\frac{1 - \sqrt{1 -
\varepsilon_s}}{m} - \pi {r_0}^2}{1-\pi
{r_0}^2}\right)}}{\left(n-1\right)\log{\left(1 + \gamma_E \psi
{r_0}^{\alpha}\right)}}\right]
\end{equation*}

\end{proof}

Based on the results of Lemma 6, we now can establish the following
theorem about the performance of Protocol 3.

\textbf{Theorem 3.} Consider the network scenario of Fig 2. To
guarantee $P_{out}^{\left(T\right)} \leq \varepsilon_t$ and
$P_{out}^{\left(S\right)} \leq \varepsilon_s$ based on the Protocol 3,
the number of eavesdroppers $m$ the network can tolerate must
satisfy the following condition.

\begin{align*}
& m \leq \frac{1- \sqrt{1-\varepsilon_s}}{\pi {r_0}^2 + \left(1- \pi
{r_0}^2\right)\omega}
\end{align*}

here,

\begin{align*}
& \omega = \left(1 + \gamma_E \psi
{r_0}^{\alpha}\right)^{-\sqrt{\frac{-\left(n-1\right)\log\left(\frac{1-\varepsilon_t}{1
- \vartheta}\right)}{\gamma_R \left(\varphi_1+\varphi_2\right)
\phi^{\alpha}}}}
\end{align*}

$\vartheta$, $\varphi_1$, $\varphi_2$, $\phi$ and $\psi$ are
defined in the same way as that in Lemma 5.

\begin{proof}

From Lemma 6, we know that to ensure the reliability requirement, we
have

\begin{align*}
\tau \leq \sqrt{\frac{-\log\left(\frac{1-\varepsilon_t}{1 -
\vartheta}\right){\phi^{-\alpha}}}{\gamma_R
{\left(n-1\right)\left(\varphi_1+\varphi_2\right)}}}
\end{align*}

and

$$\left(n-1\right)\left(1 - e^{-\tau}\right) \leq \frac{-\log{\left(\frac{1-\varepsilon_t}{1 -
\vartheta}\right)}}{\gamma_R \tau
{\phi^{\alpha}}\left(\varphi_1+\varphi_2\right)}$$

To ensure the secrecy requirement, we need

\begin{equation*}
\begin{aligned}
& m \cdot \left[\pi {r_0}^2 + \left(\frac{1}{1 + \gamma_E \psi
{r_0}^{\alpha}}\right)^{\left(n-1\right)\left(1-e^{-\tau}\right)} \left(1- \pi {r_0}^2\right) \right]\\
& \leq 1- \sqrt{1-\varepsilon_s}
\end{aligned}
\end{equation*}

Thus,

\begin{equation*}
\begin{aligned}
& m \leq \frac{1- \sqrt{1-\varepsilon_s}}{\pi {r_0}^2 +
\left(\frac{1}{1 + \gamma_E \psi
{r_0}^{\alpha}}\right)^{\left(n-1\right)\left(1-e^{-\tau}\right)} \left(1- \pi {r_0}^2\right)}\\
&\ \ \leq \frac{1- \sqrt{1-\varepsilon_s}}{\pi {r_0}^2 +
\left(\frac{1}{1 + \gamma_E \psi
{r_0}^{\alpha}}\right)^{\frac{-\log{\left(\frac{1-\varepsilon_t}{1 -
\vartheta}\right)}}{\gamma_R \tau
{\phi^{\alpha}}\left(\varphi_1+\varphi_2\right)}} \left(1- \pi
{r_0}^2\right)}
\end{aligned}
\end{equation*}

By letting $\tau$ to take its maximum value for maximum interference at
eavesdroppers, we get the following bound

\begin{align*}
& m \leq \frac{1- \sqrt{1-\varepsilon_s}}{\pi {r_0}^2 + \left(1- \pi
{r_0}^2\right)\omega}
\end{align*}

Here,

\begin{align*}
& \omega = \left(1 + \gamma_E \psi
{r_0}^{\alpha}\right)^{-\sqrt{\frac{-\left(n-1\right)\log\left(\frac{1-\varepsilon_t}{1
- \vartheta}\right)}{\gamma_R \left(\varphi_1+\varphi_2\right)
\phi^{\alpha}}}}
\end{align*}

\end{proof}

\section{Related Works}


A lot of research works have been dedicated to the implementation of physical layer
security by adopting artificial noise generation for cooperative jamming. These works can be roughly classified into two
categories depending on weather the information of eavesdroppers channels and locations is known or not.

For the case that the information of eavesdroppers channels and locations is available, many
methods can be employed to improve physical layers security by
optimizing the artificial noise generation and power control. In
case that the global channel state information is available, to
achieve the goal of maximizing the secrecy rates while minimizing the
total transmit power, a few cooperative transmission schemes have been
proposed in
\cite{IEEEhowto:Zhang1}\cite{IEEEhowto:Krikidis}\cite{IEEEhowto:Li1},
and for two-hop wireless networks the optimal transmission
strategies were presented in
\cite{IEEEhowto:Dong1}\cite{IEEEhowto:Huang}. With respect to small
networks, cooperative jamming with multiple relays and multiple
eavesdroppers and knowledge of channels and locations was considered
in \cite{IEEEhowto:Dong2}\cite{IEEEhowto:Dong3}. Even if only local
channel information rather than global channel state information is known, it was proved that the near-optimal secrecy rate can achieved by cooperative jamming schemes
\cite{IEEEhowto:Luo}\cite{IEEEhowto:Zheng}. Except channel
information, the relative locations were also considered for optimizing
cooperative jamming and power allocation to disrupt an eavesdropper
with known location \cite{IEEEhowto:Morrison}\cite{IEEEhowto:Goel}.
In addition, L. Lai et al. established the utility of user
cooperation in facilitating secure wireless communications and
proposed cooperation strategies in the additive White Gaussian Noise
(AWGN) channel \cite{IEEEhowto:Lai}, R. Negi et al. showed how
artificially generated noise can be added to the information bearing
signal to achieve secrecy in the multiple and single antenna
scenario under the constraint on total power transmitted by all
nodes \cite{IEEEhowto:Negi}. The physical layer security issues in a
two-way untrusted relay system was also investigated with friendly
jammers in \cite{IEEEhowto:Zhang2}\cite{IEEEhowto:He}. The
cooperative communications in mobile ad hoc networks was discussed in
\cite{IEEEhowto:Guan}. Effective criteria for relay and jamming node
selection were developed to ensure nonzero secrecy rate in case of
given sufficient relays in \cite{IEEEhowto:Ding1}.

For the case that the information of eavesdropper channels
and locations is unknown, the works in
\cite{IEEEhowto:Goeckel1}\cite{IEEEhowto:Goeckel2} considered the
secrecy for two-hop wireless networks, the works in
\cite{IEEEhowto:Vasudevan2}\cite{IEEEhowto:Capar1}\cite{IEEEhowto:Ding2}
considered the secrecy for large wireless networks, and the further
work in \cite{IEEEhowto:Dehghan} considered the energy efficiency
cooperative jamming strategies. These works considered how
cooperative jamming by friendly nodes can impact the security of the
network and compared it with a straightforward approach based on
multi-user diversity. They also proposed some protocols to embed
cooperative jamming techniques for protecting single links into a
large multi-hop network and explored network scaling results on
the number of eavesdroppers one network can tolerate. A.Sheikholeslami et al. explored the interference from multiple cooperative sessions
to confuse the eavesdroppers in a large wireless network \cite{IEEEhowto:Sheikholeslami}.  The cooperative relay scheme for the broadcast
channel was further investigated in \cite{IEEEhowto:Leow}\cite{IEEEhowto:Capar3}.

\section{Conclusion}

To achieve reliable and secure information transmission in a two-hop relay wireless network in presence of eavesdroppers with unknown channels and locations, several transmission protocols based on relay cooperation have been considered. In particular, theoretical analysis has been conducted to understand that under each of these protocols how many eavesdroppers one network can tolerant to meet a specified requirement on the maximum allowed secrecy outage probability and transmission outage probability.
Our results in this paper indicate that these protocols actually have different performance in terms of eavesdropper-tolerance capacity and load balance capacity among relays, and in general it is possible for us to select a proper transmission protocol according to network scenario such  that a desired trade off between the overall eavesdropper-tolerance capacity and load balance among relay nodes can be achieved.


\appendices

\section{Proof Of Lemma 1}

\begin{proof}

Notice that $P_{out}^{\left(T\right)}$ is determined as

\begin{align*}
&P_{out}^{\left(T\right)} = P\left(O_{S \rightarrow
R_{j^\ast}}^{(T)}\right)+P\left(O_{R_{j^\ast} \rightarrow
D}^{(T)}\right)\\
&\ \ \ \ \ \ \ \ \ \ -P\left(O_{S \rightarrow
R_{j^\ast}}^{(T)}\right) \cdot P\left(O_{R_{j^\ast} \rightarrow
D}^{(T)}\right)
\end{align*}

Based on the definition of transmission outage probability, we have

\begin{align*}
& P\left(O_{S
\rightarrow R_{j^\ast}}^{(T)}\right)\\
& \ \ \ \ \ = P\left(C_{S,R_{j^\ast}} \leq
\gamma_R\right)\\
& \ \ \ \ \ = P\left(\frac{E_s \cdot |h_{S,R_{j^\ast}}|^2}{\sum_{R_j
\in \mathcal {R}_1}E_s \cdot
|h_{R_j,R_{j^\ast}}|^2 + N_0/2} \leq \gamma_R\right)\\
& \ \ \ \ \ \doteq P\left(\frac{|h_{S,R_{j^\ast}}|^2}{\sum_{R_j \in
\mathcal
{R}_1}|h_{R_j,R_{j^\ast}}|^2} \leq \gamma_R\right)\\
\end{align*}

Compared to the noise generated by multiple system nodes, the
environment noise is negligible and thus is omitted here to simply
the analysis. Notice that $\mathcal {R}_1 = {\left\{j \neq j^\ast :
|h_{R_j,R_{j^\ast}}|^2 < \tau \right\}}$, then

\begin{align*}
& P\left(O_{S \rightarrow R_{j^\ast}}^{(T)}\right) \leq
P\left(\frac{|h_{S,R_{j^\ast}}|^2}{{|\mathcal {R}_1|}\tau} \leq
\gamma_R\right)\\
&  \ \ \ \ \ \ \ \ \ \ \ \ \ \ \ \ = P\left(|h_{S,R_{j^\ast}}|^2 \leq \gamma_R{|\mathcal {R}_1|}\tau\right)\\
&  \ \ \ \ \ \ \ \ \ \ \ \ \ \ \ \ \leq P\left(H^l \leq \gamma_R{|\mathcal {R}_1|}\tau\right)\\
\end{align*}

where $H^l =
min\left(\left|h_{S,R_{j^\ast}}\right|^2,\left|h_{D,R_{j^\ast}}\right|^2\right)$ is the largest random variable among the $n$ exponentially
distributed random variables
$min\left(\left|h_{S,R_j}\right|^2,\left|h_{D,R_j}\right|^2\right),
j= 1, 2, \cdots, n$.

From reference \cite{IEEEhowto:David}, we can get the distribution
function of the $min\left(|h_{S,R_j}|^2, |h_{D,R_j}|^2\right)$ for
each relay $R_j, j= 1, 2, \cdots, n$ as following,

\begin{equation*}
F_{min\left(|h_{S,R_j}|^2,
|h_{D,R_j}|^2\right)}\left(x\right)=\begin{cases} 1-e^{-2x} \ \ \ &
\text{$x > 0$}\\
0 \ \ \ &
\text{$x \leq 0$}\\
\end{cases}
\end{equation*}

From reference \cite{IEEEhowto:David}, we can also get the
distribution function of random variable $H^l$ as following,

\begin{equation*}
F_{H^l}\left(x\right)=\begin{cases} \left[1-e^{-2x}\right]^{n} \ \ \
&
\text{$x > 0$}\\
0 \ \ \ &
\text{$x \leq 0$}\\
\end{cases}
\end{equation*}

Therefore, we have

\begin{align*}
& P\left(O_{S \rightarrow R_{j^\ast}}^{(T)}\right) \leq
\left[1-e^{-2\gamma_R{|\mathcal {R}_1|}\tau}\right]^n
\end{align*}

Since there are $n - 1$ other relays except $R_{j^\ast}$, the
expected number of noise-generation nodes is given by $|\mathcal
{R}_1| =\left(n-1\right) \cdot P\left(|h_{R_j,R_{j^\ast}}|^2 <
\tau\right) = \left(n-1\right) \cdot \left(1-e^{-\tau}\right)$. Then
we have

\begin{align*}
& P\left(O_{S \rightarrow R_{j^\ast}}^{(T)}\right) \leq
\left[1-e^{-2\gamma_R\left(n-1\right)
\left(1-e^{-\tau}\right)\tau}\right]^n
\end{align*}

Employing the same method, we can get

\begin{align*}
& P\left(O_{R_{j^\ast} \rightarrow D}^{(T)}\right) \leq
\left[1-e^{-2\gamma_R\left(n-1\right)
\left(1-e^{-\tau}\right)\tau}\right]^n
\end{align*}

Thus, we have

\begin{align*}
&P_{out}^{\left(T\right)} \leq
2\left[1-e^{-2\gamma_R\left(n-1\right)
\left(1-e^{-\tau}\right)\tau}\right]^n\\
&\ \ \ \ \ \ \ \ \ \ -\left[1-e^{-2\gamma_R\left(n-1\right)
\left(1-e^{-\tau}\right)\tau}\right]^{2n}
\end{align*}

Similarly, notice that $P_{out}^{\left(S\right)}$ is given by

\begin{align*}
&P_{out}^{\left(S\right)} = P\left(O_{S \rightarrow
R_{j^\ast}}^{(S)}\right)+P\left(O_{R_{j^\ast} \rightarrow
D}^{(S)}\right)\\
&\ \ \ \ \ \ \ \ \ -P\left(O_{S \rightarrow R_{j^\ast}}^{(S)}\right)
\cdot P\left(O_{R_{j^\ast} \rightarrow D}^{(S)}\right)
\end{align*}

According to the definition of secrecy outage probability, we know
that

\begin{align*}
&P\left(O_{S \rightarrow R_{j^\ast}}^{(S)}\right) =
P\left(\bigcup_{i=1}^{m}\left\{C_{S,E_i} \geq
\gamma_E\right\}\right)
\end{align*}

Thus, we have

\begin{align*}
&P\left(O_{S \rightarrow R_{j^\ast}}^{(S)}\right) \leq
\sum_{i=1}^{m}P\left(C_{S,E_i} \geq \gamma_E\right)
\end{align*}

Based on the Markov inequality,

\begin{align*}
& P\left(C_{S,E_i} \geq \gamma_E\right)\\
& \ \ \ \ \ \leq P\left(\frac{E_s \cdot |h_{S,E_i}|^2}{\sum_{R_j \in
\mathcal {R}_1}E_s \cdot |h_{R_j,E_i}|^2} \geq
\gamma_E\right)\\
& \ \ \ \ \ = E_{\left\{h_{R_j,E_i}, j=0,1,\cdots,n+mp,j \neq
j^{\ast}\right\},\mathcal {R}_1}\\
& \ \ \ \ \ \ \ \ \ \left[P\left(|h_{S,E_i}|^2 > \gamma_E
\cdot \sum_{R_j \in \mathcal {R}_1}|h_{R_j,E_i}|^2\right)\right]\\
& \ \ \ \ \ \leq E_{\mathcal {R}_1}\left[\prod_{R_j \in \mathcal
{R}_1}
E_{h_{R_j,E_i}}\left[e^{-\gamma_E|h_{R_j,E_i}|^2}\right]\right]\\
& \ \ \ \ \ = E_{\mathcal
{R}_1}\left[\left(\frac{1}{1+\gamma_E}\right)^{|\mathcal
{R}_1|}\right]
\end{align*}

Therefore,

$$P\left(O_{S \rightarrow R_{j^\ast}}^{(S)}\right) \leq
\sum_{i=1}^{m}\left(\frac{1}{1+\gamma_E}\right)^{|\mathcal {R}_1|} =
m \cdot \left(\frac{1}{1+\gamma_E}\right)^{|\mathcal {R}_1|}$$

Employing the same method, we can get

$$P\left(O_{R_{j^\ast} \rightarrow D}^{(S)}\right) \leq
m \cdot \left(\frac{1}{1+\gamma_E}\right)^{|\mathcal {R}_2|}$$

Since the expected number of noise-generation nodes is given by
$|\mathcal {R}_1| = |\mathcal {R}_2| = \left(n-1\right) \cdot
\left(1-e^{-\tau}\right)$, thus, we can get

\begin{align*}
&P_{out}^{\left(S\right)} \leq 2m \cdot
\left(\frac{1}{1+\gamma_E}\right)^{\left(n-1\right) \cdot
\left(1-e^{-\tau}\right)}\\
&\ \ \ \ \ \ \ \ \ -\left[m \cdot
\left(\frac{1}{1+\gamma_E}\right)^{\left(n-1\right) \cdot
\left(1-e^{-\tau}\right)}\right]^2
\end{align*}

\end{proof}

\section{Proof Of Lemma 3}

\begin{proof}

Similar to the proof of Lemma 1, we notice that
$P_{out}^{\left(T\right)}$ is determined as

\begin{align*}
&P_{out}^{\left(T\right)} = P\left(O_{S \rightarrow
R_{j^\ast}}^{(T)}\right)+P\left(O_{R_{j^\ast} \rightarrow
D}^{(T)}\right)\\
&\ \ \ \ \ \ \ \ \ \ -P\left(O_{S \rightarrow
R_{j^\ast}}^{(T)}\right) \cdot P\left(O_{R_{j^\ast} \rightarrow
D}^{(T)}\right)
\end{align*}

Based on the definition of transmission outage probability, we have

\begin{align*}
& P\left(O_{S \rightarrow R_{j^\ast}}^{(T)}\right) =
P\left(C_{S,R_{j^\ast}} \leq
\gamma_R\right)\\
& \ \ \ \ \ \ \ \ \ \ \ \ \ \ \ \ \ \leq
P\left(\frac{|h_{S,R_{j^\ast}}|^2}{{|\mathcal {R}_1|}\tau} \leq
\gamma_R\right)\\
& \ \ \ \ \ \ \ \ \ \ \ \ \ \ \ \ \ = P\left(|h_{S,R_{j^\ast}}|^2 \leq \gamma_R{|\mathcal {R}_1|}\tau\right)\\
& \ \ \ \ \ \ \ \ \ \ \ \ \ \ \ \ \ = 1 - e^{-\gamma_R{|\mathcal
{R}_1|}\tau}
\end{align*}

Here $\mathcal {R}_1 = {\left\{j \neq j^\ast :
|h_{R_j,R_{j^\ast}}|^2 < \tau \right\}}$. Since the expected number
of noise-generation nodes is given by $|\mathcal {R}_1| =
\left(n-1\right) \cdot \left(1-e^{-\tau}\right)$. Then we have

\begin{align*}
&P\left(O_{S \rightarrow R_{j^\ast}}^{(T)}\right) \leq 1 -
e^{-\gamma_R\left(n -1\right)\left(1-e^{-\tau}\right)\tau}
\end{align*}

Employing the same method, we can get

\begin{align*}
&P\left(O_{R_{j^\ast} \rightarrow D}^{(T)}\right) \leq 1 -
e^{-\gamma_R\left(n -1\right)\left(1-e^{-\tau}\right)\tau}
\end{align*}

Thus, we have

\begin{align*}
&P_{out}^{\left(T\right)} \leq 2\left[1 -
e^{-\gamma_R\left(n -1\right)\left(1-e^{-\tau}\right)\tau}\right]\\
&\ \ \ \ \ \ \ \ \ \ -\left[1 - e^{-\gamma_R\left(n
-1\right)\left(1-e^{-\tau}\right)\tau}\right]^2
\end{align*}

Notice that the eavesdropper model of Protocol 1 is the same as that
of Protocol 2, the method for ensuring secrecy is identical to that
of in Lemma 1. Thus, we can see that the secrecy outage probability
of Protocol 1 and Protocol 2 is the same, that is,

\begin{align*}
&P_{out}^{\left(S\right)} \leq 2m \cdot
\left(\frac{1}{1+\gamma_E}\right)^{\left(n-1\right) \cdot
\left(1-e^{-\tau}\right)}\\
&\ \ \ \ \ \ \ \ \ -\left[m \cdot
\left(\frac{1}{1+\gamma_E}\right)^{\left(n-1\right) \cdot
\left(1-e^{-\tau}\right)}\right]^2
\end{align*}

\end{proof}

\section{Proof Of Lemma 5}

\begin{proof}

Notice that two ways leading to transmission outage are: 1) there are
no candidate relays in the relay selection region; 2) the SINR at
the selected relay or the destination is less than $\gamma_R$. Let
$A_1$ be the event that there is at least one relay in the relay
selection region, and $A_2$ be the event that there are no relays in
the relay selection region. We have

\begin{align*}
&P_{out}^{\left(T\right)} = P_{out|A_1}^{\left(T\right)}P(A_1) +
P_{out|A_2}^{\left(T\right)}P(A_2)
\end{align*}

Since the relay is uniformly distributed, the number of candidate
relays is a binomial distribution $\bigg(n,
\left(1-2a\right)\left(1-2b\right)\bigg)$. We have

$$P(A_1) = 1 - \vartheta$$

and

$$P(A_2) = \vartheta$$

where $\vartheta =
\bigg[1-\left(1-2a\right)\left(1-2b\right)\bigg]^n$. When event
$A_2$ happens, no relay is available. Then

\begin{align*}
&P_{out|A_2}^{\left(T\right)} = 1
\end{align*}

Thus, we have

\begin{align*}
&P_{out}^{\left(T\right)} = P_{out|A_1}^{\left(T\right)}\left(1 -
\vartheta\right)+ 1 \cdot \vartheta
\end{align*}

Notice that $P_{out|A_1}^{\left(T\right)}$ is determined as

\begin{align*}
&P_{out|A_1}^{\left(T\right)} = P\left(O_{S \rightarrow
R_{j^\ast}}^{(T)}\bigg|A_1\right)+P\left(O_{R_{j^\ast} \rightarrow
D}^{(T)}\bigg|A_1\right)\\
&\ \ \ \ \ \ \ \ \ \ \ \ -P\left(O_{S \rightarrow
R_{j^\ast}}^{(T)}\bigg|A_1\right) \cdot P\left(O_{R_{j^\ast}
\rightarrow D}^{(T)}\bigg|A_1\right)
\end{align*}

Based on the definition of transmission outage probability, we have

\begin{align*}
& P\left(O_{S
\rightarrow R_{j^\ast}}^{(T)}\bigg|A_1\right)\\
& \ \ \ \ \ = P\left(C_{S,R_{j^\ast}} \leq
\gamma_R\bigg|A_1\right)\\
& \ \ \ \ \ = P\left(\frac{E_s \cdot
\frac{|h_{S,R_{j^\ast}}|^2}{d_{S,R_{j^\ast}}^{\alpha}}}{\sum_{R_j
\in \mathcal {R}_1}E_s \cdot
\frac{|h_{R_j,R_{j^\ast}}|^2}{d_{R_j,R_{j^\ast}}^{\alpha}} + \frac{N_0}{2}} \leq \gamma_R\bigg|A_1\right)\\
& \ \ \ \ \ \doteq P\left(\frac{
\frac{|h_{S,R_{j^\ast}}|^2}{d_{S,R_{j^\ast}}^{\alpha}}}{\sum_{R_j
\in \mathcal {R}_1}
\frac{|h_{R_j,R_{j^\ast}}|^2}{d_{R_j,R_{j^\ast}}^{\alpha}}} \leq \gamma_R\bigg|A_1\right)\\
\end{align*}

Compared to the noise generated by multiple system nodes, the
environment noise is negligible and thus is omitted here to simply
the analysis. Notice that $\mathcal {R}_1 = {\left\{j \neq j^\ast :
|h_{R_j,R_{j^\ast}}|^2 < \tau \right\}}$, then

\begin{align*}
& P\left(O_{S \rightarrow R_{j^\ast}}^{(T)}\bigg|A_1\right) \leq
P\left(\frac{|h_{S,R_{j^\ast}}|^2 d_{S,R_{j^\ast}}^{-\alpha}} {
\sum_{R_j \in \mathcal {R}_1} \tau d_{R_j,R_{j^\ast}}^{-\alpha}}
\leq \gamma_R\bigg|A_1\right)\\
\end{align*}

As shown in Fig 2 that by assuming the coordinate of $R_j$ as $\left(x, y\right)$, we can see that the number of noise generating nodes in square $\left[x,x+dx\right] \times \left[y, y+dy\right]$ will be $\left(n-1\right)\left(1
- e^{-\tau}\right)dxdy$. Then, we have

\begin{align*}
& \sum_{R_j \in \mathcal {R}_1}
\frac{\tau}{d_{R_j,R_{j^\ast}}^{\alpha}} \\
&\ \ \ \ \ \ \ \ = \int_0^1\int_0^1\frac{\tau
\left(n-1\right)\left(1
- e^{-\tau}\right)}{\left[\left(x - x_{R_{j^{\ast}}}\right)^2 + \left(y - y_{R_{j^{\ast}}}\right)^2 \right]^{\frac{\alpha}{2}}}dxdy\\
\end{align*}

where $\left(x_{R_{j^{\ast}}}, y_{R_{j^{\ast}}}\right)$ is the
coordinate of the selected relay $R_{j^{\ast}}$, $x_{R_{j^{\ast}}}
\in \left[a, 1-a\right], y_{R_{j^{\ast}}} \in \left[b, 1-b\right]$
and $a \in \left[0, 0.5\right], b \in \left[0, 0.5\right]$.

Notice that within the network area, where relays are uniformly distributed, the worst case location for the
selected relay $R_{j^{\ast}}$ is the point $\left(0.5,
0.5\right)$, at which the interference from the noise
generating nodes is the largest; whereas, the best case location for the selected
relay $R_{j^{\ast}}$ is the four corner points $(a,b), (a,1-b), (1-a,b)$ and
$(1-a,1-b)$ of the relay selection, where the interference from the noise
generating nodes is the smallest. By considering the worst case location for the selected relay $R_{j^{\ast}}$, we have

$$P\left(O_{S \rightarrow R_{j^\ast}}^{(T)}\bigg|A_1\right) \leq
P\left(\frac{|h_{S,R_{j^\ast}}|^2 d_{S,R_{j^\ast}}^{-\alpha}} { \tau
\left(n-1\right)\left(1 - e^{-\tau}\right) \varphi_1 } \leq
\gamma_R\bigg|A_1\right)$$

Here

$$\varphi_1 =  \int_0^1\int_0^1\frac{1}{\left[\left(x -
0.5\right)^2 + \left(y -
0.5\right)^2\right]^{\frac{\alpha}{2}}}dxdy$$

Due to $a \leq d_{S,R_{j^\ast}} \leq \sqrt{(1-a)^2+(0.5-b)^2}$, and
let $\phi = \sqrt{(1-a)^2+(0.5-b)^2}$, then

\begin{align*}
& P\left(O_{S \rightarrow R_{j^\ast}}^{(T)}\bigg|A_1\right) \\
&\ \ \ \leq P\left(\frac{|h_{S,R_{j^\ast}}|^2 \phi^{-\alpha}} {\tau
\left(n-1\right)\left(1 - e^{-\tau}\right) \varphi_1 } \leq \gamma_R\bigg|A_1\right)\\
&\ \ \ = P\left(|h_{S,R_{j^\ast}}|^2 \leq \frac{\gamma_R {\tau
\left(n-1\right)\left(1 - e^{-\tau}\right)
\varphi_1}}{\phi^{-\alpha}}\bigg|A_1\right)\\
&\ \ \ = 1-e^{-\frac{\gamma_R {\tau \left(n-1\right)\left(1 -
e^{-\tau}\right) \varphi_1}}{\phi^{-\alpha}}}
\end{align*}

Employing the same method, we can get

\begin{align*}
& P\left(O_{R_{j^\ast} \rightarrow D}^{(T)}\bigg|A_1\right) \leq
1-e^{-\frac{\gamma_R {\tau \left(n-1\right)\left(1 -
e^{-\tau}\right) \varphi_2}}{\phi^{-\alpha}}}
\end{align*}

here,

$$\varphi_2 =  \int_0^1\int_0^1\frac{1}{\left[\left(x -
1\right)^2 + \left(y - 0.5\right)^2\right]^{\frac{\alpha}{2}}}dxdy$$

Then, we have

\begin{align*}
&P_{out|A_1}^{\left(T\right)} \leq 1-e^{-\frac{\gamma_R {\tau
\left(n-1\right)\left(1 -
e^{-\tau}\right)}}{\phi^{-\alpha}}\left(\varphi_1+\varphi_2\right)}
\end{align*}

Thus, we have

\begin{equation*}
\begin{aligned}
&P_{out}^{\left(T\right)} \leq \left[1-e^{-\frac{\gamma_R {\tau
\left(n-1\right)\left(1 -
e^{-\tau}\right)}}{\phi^{-\alpha}}\left(\varphi_1+\varphi_2\right)}\right]\left(1
- \vartheta\right)+ 1 \cdot \vartheta
\end{aligned}
\end{equation*}

Notice that $P_{out}^{\left(S\right)}$ is given by

\begin{align*}
&P_{out}^{\left(S\right)} = P\left(O_{S \rightarrow
R_{j^\ast}}^{(S)}\right)+P\left(O_{R_{j^\ast} \rightarrow
D}^{(S)}\right)\\
&\ \ \ \ \ \ \ \ \ -P\left(O_{S \rightarrow R_{j^\ast}}^{(S)}\right)
\cdot P\left(O_{R_{j^\ast} \rightarrow D}^{(S)}\right)
\end{align*}

According to the definition of secrecy outage probability, we know
that

\begin{align*}
&P\left(O_{S \rightarrow R_{j^\ast}}^{(S)}\right) =
P\left(\bigcup_{i=1}^{m}\left\{C_{S,E_i} \geq
\gamma_E\right\}\right)
\end{align*}

Thus, we have

\begin{align*}
&P\left(O_{S \rightarrow R_{j^\ast}}^{(S)}\right) \leq
\sum_{i=1}^{m}P\left(C_{S,E_i} \geq \gamma_E\right)
\end{align*}

Based on the definition of $r_0$, we denote by $G_1^{(i)}$ the event that the distance between $E_i$ and the source
is less than $r_0$, and denote by $G_2^{(i)}$ the event that distance between $E_i$
and the source is lager than or equal to $r_0$. We have

\begin{equation*}
\begin{aligned}
&P\left(C_{S,E_i} \geq \gamma_E\right) \\
&\ \ \ \ =
P\left(C_{S,E_i} \geq \gamma_E\bigg|G_1^{(i)}\right)P\left(G_1^{(i)}\right)\\
&\ \ \ \ \ \ \ \ +
P\left(C_{S,E_i} \geq \gamma_E\bigg|G_2^{(i)}\right)P\left(G_2^{(i)}\right)\\
&\ \ \ \ \leq 1 \cdot \frac{1}{2} \pi {r_0}^2 +
P\left(C_{S,E_i} \geq \gamma_E\bigg|G_2^{(i)}\right)\left(1- \frac{1}{2} \pi {r_0}^2\right)\\
\end{aligned}
\end{equation*}

of which

\begin{equation*}
\begin{aligned}
&P\left(C_{S,E_i} \geq \gamma_E\bigg|G_2^{(i)}\right)\\
&\ \ \leq P\left(\frac{|h_{S,E_i}|^2 {r_0}^{-\alpha}}{\Gamma
\int_0^1\int_0^1\frac{1}{\left[\left(x - x_{E_i}\right)^2 + \left(y
- y_{E_i}\right)^2\right]^{\frac{\alpha}{2}}}dxdy}\geq
\gamma_E\bigg|G_2^{(i)}\right)
\end{aligned}
\end{equation*}

where $\left(x_{E_i}, y_{E_i}\right)$ is the coordinate of the
eavesdropper $E_i$. $\Gamma$ is the sum of
$\left(n-1\right)\left(1-e^{-\tau}\right)$ independent exponential
random variables.

From Fig 2 we know that the largest interference at eavesdropper $E_i$ happens
when $E_i$ is located at the point $(0.5, 0.5)$, while the smallest interference at $E_i$ happens it is located at the four corners of the
network region. By considering  the smallest interference at eavesdroppers, we then have

\begin{align*}
&P\left(C_{S,E_i} \geq \gamma_E\bigg|G_2^{(i)}\right)\\
&\ \ \ \ \ \ \leq P\left(\frac{|h_{S,E_i}|^2 {r_0}^{-\alpha}}{\Gamma
\psi }\geq
\gamma_E\right)\\
&\ \ \ \ \ \ = P\left(|h_{S,E_i}|^2 \geq \Gamma \gamma_E \cdot \psi
\cdot {r_0}^{\alpha} \right)
\end{align*}

here

$$\psi =
\int_0^1\int_0^1\frac{1}{\left(x^2 +
y^2\right)^{\frac{\alpha}{2}}}dxdy$$

Based on the Markov inequality,

\begin{align*}
& P\left(C_{S,E_i} \geq \gamma_E\bigg|G_2^{(i)}\right)\\
& \ \ \ \ \ \leq E_\Gamma \left[e^{-\Gamma \gamma_E \psi
{r_0}^{\alpha}}\right]\\
& \ \ \ \ \ = \left(\frac{1}{1 + \gamma_E \psi
{r_0}^{\alpha}}\right)^{\left(n-1\right)\left(1-e^{-\tau}\right)}\\
\end{align*}

Then, we have

\begin{equation*}
\begin{aligned}
& P\left(C_{S,E_i} \geq \gamma_E\right) \\
&\leq \frac{1}{2} \pi {r_0}^2 +\left(\frac{1}{1 + \gamma_E \psi
{r_0}^{\alpha}}\right)^{\left(n-1\right)\left(1-e^{-\tau}\right)}\left(1- \frac{1}{2} \pi {r_0}^2\right)\\
\end{aligned}
\end{equation*}

Employee the same method, we have

\begin{equation*}
\begin{aligned}
& P\left(C_{R_{j^\ast},E_i} \geq \gamma_E\right) \\
&\leq \pi {r_0}^2 + \left(\frac{1}{1 + \gamma_E \psi
{r_0}^{\alpha}}\right)^{\left(n-1\right)\left(1-e^{-\tau}\right)}\left(1- \pi {r_0}^2\right)\\
\end{aligned}
\end{equation*}

Notice that

\begin{equation*}
\begin{aligned}
&\frac{1}{2} \pi {r_0}^2 +\left(\frac{1}{1 + \gamma_E \psi
{r_0}^{\alpha}}\right)^{\left(n-1\right)\left(1-e^{-\tau}\right)}\left(1- \frac{1}{2} \pi {r_0}^2\right)\\
&= \pi {r_0}^2 + \left(\frac{1}{1 + \gamma_E \psi
{r_0}^{\alpha}}\right)^{\left(n-1\right)\left(1-e^{-\tau}\right)}\left(1- \pi {r_0}^2\right)\\
&\ \ \ \ -\frac{1}{2} \pi {r_0}^2 \left[1-\left(\frac{1}{1 +
\gamma_E \psi
{r_0}^{\alpha}}\right)^{\left(n-1\right)\left(1-e^{-\tau}\right)}\right]\\
& \leq \pi {r_0}^2 + \left(\frac{1}{1 + \gamma_E \psi
{r_0}^{\alpha}}\right)^{\left(n-1\right)\left(1-e^{-\tau}\right)}\left(1-
\pi {r_0}^2\right)
\end{aligned}
\end{equation*}

Therefore

\begin{equation*}
\begin{aligned}
&P\left(O_{S \rightarrow R_{j^\ast}}^{(S)}\right) \leq P\left(O_{R_{j^\ast} \rightarrow D}^{(S)}\right)  \\
&\leq m \left[\pi {r_0}^2 + \left(\frac{1}{1 + \gamma_E \psi
{r_0}^{\alpha}}\right)^{\left(n-1\right)\left(1-e^{-\tau}\right)}\left(1- \pi {r_0}^2\right)\right]\\
\end{aligned}
\end{equation*}

Then, we have

\begin{equation*}
\begin{aligned}
&P_{out}^{\left(S\right)} \leq 2 m \left[\pi {r_0}^2 +
\left(\frac{1}{1 + \gamma_E \psi
{r_0}^{\alpha}}\right)^{\left(n-1\right)\left(1-e^{-\tau}\right)}\left(1- \pi {r_0}^2\right)\right]\\
&\ \ -\left[m \left(\pi {r_0}^2 + \left(\frac{1}{1 + \gamma_E \psi
{r_0}^{\alpha}}\right)^{\left(n-1\right)\left(1-e^{-\tau}\right)}\left(1-
\pi {r_0}^2\right)\right)\right]^2
\end{aligned}
\end{equation*}

\end{proof}



\ifCLASSOPTIONcaptionsoff
\newpage
\fi

\end{document}